\documentclass[aip,apl,reprint,twocolumn,superscriptaddress]{revtex4-1}
\usepackage{amsmath,amssymb,graphicx,epstopdf}

\begin{document}
\title{Optical amplifier based on guided polaritons in GaN and ZnO}

\author{D. D. Solnyshkov}
\affiliation{LASMEA, Clermont Universit\'e, Universit\'e Blaise Pascal, CNRS, Clermont-Ferrand, France}

\author{H. Ter\c{c}as}
\affiliation{LASMEA, Clermont Universit\'e, Universit\'e Blaise Pascal, CNRS, Clermont-Ferrand, France}

\author{G. Malpuech}
\affiliation{LASMEA, Clermont Universit\'e, Universit\'e Blaise Pascal, CNRS, Clermont-Ferrand, France}

\date{\today}

\begin{abstract}
We propose a scheme of an optical amplifier based on GaN and ZnO waveguides operating in the regime of strong coupling between photonic modes and excitonic resonances. Amplification of the guided exciton-polaritons is obtained by stimulated scattering from the excitonic reservoir, which is found to be fast enough compared with the large velocity of the guided polariton modes. We analyze the device parameters at different temperatures. We find that an 80 $\mu$m-long amplifier can provide a gain of 10dB at room temperature, being supplied by 5 mA current in the $cw$ regime.
\end{abstract}
\maketitle

There is presently a large activity aiming to realize low consumption, compact size all-optical devices, which could replace electronics for some tasks. Recent developments based on the use of photonic crystals and on different types of non-linearities (quantum dots, optical index modulation) show good promises \cite{Ambs2010,Nozaki2010}. Industrial implementation of some solutions looks plausible within the next decade. In that framework, exciton-polaritons, due to their high nonlinear response, have important advantages. Cavity polaritons\cite{CavityPolaritons2003} are two-dimensional quasi-particles arising from the coupling of excitons present in the cavity (either bulk or quantum well excitons) and the fundamental mode of a Fabry-Perot resonator. The first device proposed in this field is the so-called polariton laser based on polariton condensation\cite{Imamoglu1996}. It does not require the achievement of population inversion and has a very low threshold. Polariton lasing has been observed in optically pumped GaN and ZnO-based cavities at low and room temperature  \cite{Christopoulos2007,Baumberg2008,Feng2013,Trichet2013}. Electrically pumped polariton lasers \cite{polelec,patent} running at low \cite{Schneider,bath1} and room temperature \cite{bath2} have also been recently reported. 

In the past years, several other types of polariton devices, such as circuits, switches, and transistors have been designed theoretically, and some of them implemented. Most realized schemes are based on a polariton flow having a few meVs of kinetic energy, propagating at a few percents of the speed of light. This flow acts as the useful signal and its propagation can be controlled by an optical signal from an external laser which serves as a gate. In Refs.\cite{Amo2010,Ballarini2013}, the creation of the flow from a pumped area is triggered by a resonant pulse which allows the transition to the upper state of a bistable oscillator. In Ref.\cite{Gao2012}, the flow is created by a resonant laser under oblique incidence in a wire microcavity, while a second laser creates an excitonic cloud in an area of a few $\mu$m. This cloud acts on the excitonic part of the propagating polaritons and creates a potential barrier which stops the flow (transistor off). The repetition rate is about 10 GHz using a non-resonant control laser, but can go up to 500 GHz using a resonant control. Polaritonic Resonant tunneling transistors \cite{Nguyen2013} and Mach-Zehnder interferometers  \cite{Sturm2014} have been implemented as well. Such devices are based on propagative cavity modes with a short/moderate life time hardly exceeding 100 ps. These modes are slow light, with a propagation velocity of the order of $0.3\%$ of $c$.

The optical amplifiers are the most useful of all the devices used in any optical communication scheme. The optical waveguides are ubiquitous in modern communication technologies. At telecom wavelengths, the optical signal can propagate for very large distances, but because of the finite extinction it has to be re-amplified regularly. The efficiency and rapidity of such amplification are important for practical purposes. Many proposals were made in scientific papers and patents for semiconductor optical amplifiers \cite{Sakuda1988,Akiyama2005}. Such amplifiers are based on different material systems and active medium configurations (3D-0D) \cite{Yamamoto1989,Capua2014,Sugawara2004}. It is important to note that their typical lengths are on the mm scale. Many efforts are devoted to the decrease of the operation period of these devices \cite{Schubert2004}. 
Re-amplification is even more crucial for other frequency domains, where the propagation distances are much shorter.

In this work, we theoretically study an optical amplifier, based on a waveguide in the regime of strong coupling \cite{Walker2013,Oder2002}. This is really an analogue of a semiconductor optical amplifier, where the beam is re-amplified by a gain medium, but based on a different mechanism. The excitonic reservoir providing the gain can be created by optical or by electrical pumping. The waveguide scheme allows much easier realization of a contacted PN junction.  The key question we wish to reply is if the scattering rate from an excitonic reservoir to a guided polariton mode, which plays the role of the gain in a polariton system, can be fast enough compared with the large propagation velocity in a waveguide. The stimulated scattering into the propagating polariton mode has been recently demonstrated in wire cavities \cite{Wertz2012}, but with slowly propagating modes. In strongly coupled waveguides, the modes can be 50 times faster. We calculate the typical device size to get a substantial re-amplification ratio. We propose to consider amplifiers based on two different wide-bandgap semiconductors: GaN and ZnO. The high exciton binding energy and oscillator strength in this material allow to exploit the excitonic effects, such as the strong exciton-photon coupling, at room temperature \cite{Malpuech2002,Zamfirescu,polelec,patent,Christopoulos2007,Baumberg2008,Feng2013,bath2}. The wide-bandgap semiconductors are active in the blue part of the optical spectrum, far away from telecom wavelengths. The designed amplifier cannot be used in telecom lines, but it can be important for short-range applications, such as optical computers and optical buses in mixed computers which will naturally tend to develop in the future.

\begin{figure}[t]
\begin{center}
\includegraphics[width=0.99\linewidth]{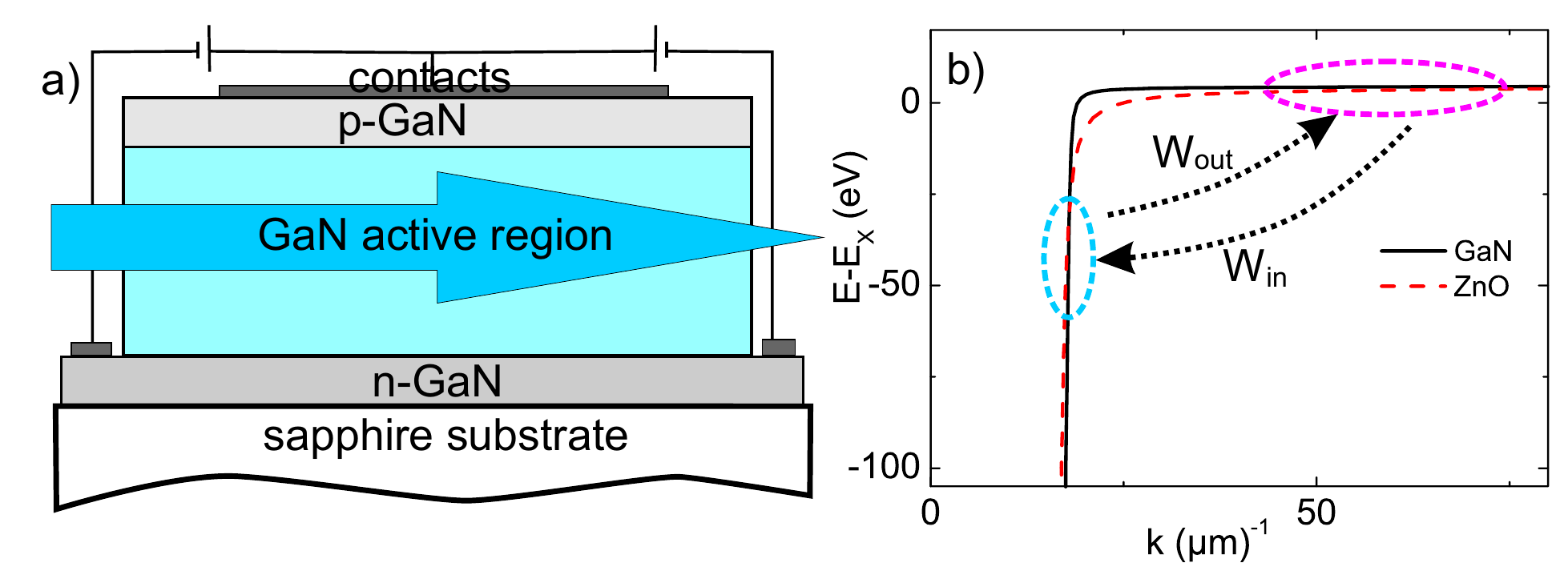}
\caption {\label{fig1} a) The scheme of the polariton amplifier. The two edges of the waveguide can be doped for electrical injection. b) The dispersion of the waveguide  in the strong coupling regime (\eqref{dispersion},\eqref{disppol}, showing the region of amplification (red) and the reservoir (blue), as well as the scattering rates $W_{in}$,$W_{out}$}
\end{center}
\end{figure}

The scheme of the device is shown in Fig. 1a). The waveguide represents a GaN slab with doped regions and metallic contacts for electrical injection. The simple configuration we propose could be further enhanced to avoid the losses in the metallic contacts. We consider the waveguide thickness of $d_{GaN}$=65~nm and $d_{ZnO}$=52~nm for monomode operation.

The signal (a gaussian wavepacket) is injected into the guided mode on the left and propagates to the right. The amplification coefficient is defined as the ratio of intensities at the input and at the output. Fig. 1b) shows the polariton dispersion in the strong coupling regime. It is obtained from the coupling between the exciton mode $E_X$ (almost flat at this scale) and the waveguide dispersion given by the equation 
\begin{equation}
\label{dispersion}
\tan ^2 \left( {\frac{d}
{2}\sqrt {\left( {\frac{\omega }
{{c_1 }}} \right)^2  - \beta ^2 }  - \frac{{m\pi }}
{2}} \right) = \frac{{\beta ^2  - {{\omega ^2 } \mathord{\left/
 {\vphantom {{\omega ^2 } {c_2^2 }}} \right.
 \kern-\nulldelimiterspace} {c_2^2 }}}}
{{{{\omega ^2 } \mathord{\left/
 {\vphantom {{\omega ^2 } {c_1^2  - \beta ^2 }}} \right.
 \kern-\nulldelimiterspace} {c_1^2  - \beta ^2 }}}}
\end{equation}
where $\beta=n_{eff}\omega/c$ is the propagation constant, $\omega$ is the mode frequency, $c_1,c_2,c$ are the light velocities in the core, cladding, and vacuum, respectively, and $m=0$ is the mode number (we consider a single-mode waveguide). The dispersion of the lower polariton branch was calculated taking into account a Rabi splitting of 70 meV for bulk GaN active region and 160 meV for the ZnO:
\begin{equation}
\label{disppol}
E_{pol}=\frac{E_X+\hbar\omega}{2}-\sqrt{\left(\frac{E_X-\hbar\omega}{2}\right)^2+\hbar^2\Omega^2}
\end{equation}
 This dispersion is shown in Fig. 2b for GaN (black) and ZnO (red), with the exciton energy taken as the zero reference in both cases.

 The exciton reservoir is marked with magenta dashed ellipse, and the polariton mode suitable for amplification is shown with blue dashed lines. For photons with lower energies the scattering from the reservoir will be strongly suppressed, while for energies closer to the reservoir the propagation will be too slow, and the temperature depletion towards the reservoir will play a strong role. It is therefore important to find the optimal energy for sending the signal to be amplified.
 
We describe the relaxation of exciton-polaritons from the reservoir towards the guided mode using the semi-classical Boltzmann equations for a bulk waveguide according to the general procedure described in Ref.\cite{CavityPolaritons2003,SolnyshkovJAP2008}, taking into account the effective decay rate due to the nonzero group velocity $v_g=\hbar^{-1}\partial E/\partial k$ of the guided mode.  

\begin{equation}
\begin{split}
\frac{dn_\textbf{k}}{dt}=&P_\textbf{k}-\Gamma_\textbf{k} n_\textbf{k}- n_\textbf{k}\sum_{\textbf{k}'}W_{\textbf{k}\rightarrow \textbf{k}'}(n_{\textbf{k}'}+1)+\\
&+(n_\textbf{k}+1)\sum_{\textbf{k}'}W_{\textbf{k}'\rightarrow \textbf{k}}n_\textbf{k}'\,. \label{boltzmann}
\end{split}
 \end{equation}
Here, $n_k$ is the polariton distribution function, $\Gamma_k=v_g/\Delta x$ is the effective escape rate of the particles from a spatial cell of a size $\Delta x=10~\mu$m, $P_{\textbf{k}}=P_{0}\exp{(-(E_{\textbf{k}}-E_{X})/k_{B}T)}$ is the pumping, acting in the excitonic reservoir.$W_{\textbf{k}\rightarrow \textbf{k}'}$ are the semiclassical Boltzmann rates associated with the exciton-phonon scattering (both acoustic and LO) and exciton-exciton scattering. We are therefore studying what happens at the left edge of the amplifier, in order to find the steady state regime and the optimal energies, where the amplification will be strongest. The three excitonic resonances present in GaN and ZnO are assimilated to a single effective exciton resonance. The steady state regime should provide a polariton population close to the threshold value $n_{pol}\approx 1$ for the range of energies in question.

The results of the simulations are shown in Fig. 2a), demonstrating the net scattering rate as a function of energy at 300 K. This scattering rate will provide the amplification of the propagating signal if it exceeds the losses. A peak is clearly observed at an energy of about 35 meV below the exciton energy (chosen as the zero reference) for GaN and about 40 meV for ZnO. This maximum appears because of the interplay of the excitonic fraction (decreasing with energy) and the thermal depletion (increasing when approaching the reservoir). The optimal regime for the amplifier is just below the lasing threshold: $3.1\times 10^{16}$ and $1.2\times 10^{16}$ cm$^{-3}$ for GaN and ZnO correspondingly. The injection current corresponding to this threshold density is of the order of 5~mA. The values of the net scattering rate are lower for ZnO, because of the lower group velocity (due to the higher Rabi splitting) and thus lower effective escape rates of the particles (compensating this net scattering rate). The LO phonons play an important role in ZnO, because their energy $E_{\textrm{LO}}=72$~meV is comparable with the Rabi splitting. The energy of the optimal mode -40~meV corresponds to the 1st phonon replica of the reservoir $E_{\textrm{1LO}}=3/2k_B T -E_{\textrm{LO}}$, which makes possible direct scattering from the reservoir into the polariton mode \cite{Orosz2012}.

In order to simulate the spatial effects, we have rewritten the semi-classical Boltzmann equations with a spatial resolution, but reducing the reservoir to a single effective "state" \cite{Galbiati2012} with effective scattering rates in and out of the guided mode $W_{in}$ and $W_{out}$ found from the previous full Boltzmann simulations.

\begin{eqnarray}
\label{propag}
\frac{{dN}}
{{dt}} &=& W_{in} N_r^2  + \left( {W_{in} N_r  - W_{out} } \right)N_r N - v\frac{{dN}}
{{dx}} \\
\frac{{dN_r }}
{{dt}} &=&  - W_{in} N_r^2  - \left( {W_{in} N_r  - W_{out} } \right)N_r N
\end{eqnarray}

This pair of equations is equivalent to the pair of rate equations usually written to describe the light propagation in a semiconductor optical amplifier \cite{Haridim2011,Agrawal1989}, the difference being that instead of dealing with carriers and photons, we have an exciton reservoir and a polariton mode, and the mechanism of scattering is different. In usual semiconductor amplifiers, an electron and a hole disappear, generating a photon, while in polariton systems two excitons in the reservoir scatter on each other, one going into the polariton mode and the other gaining energy and then relaxing backwards with phonons.

In order to obtain analytical estimate of the amplification efficiency, we write the following equation only for the propagating polariton mode, assuming that the signal is weak and that the reservoir population does not change, which allows to include it into the scattering rates $W_{in,out}'$: 

\[
\frac{{dN}}
{{dt}} = W_{in}'  + \left( {W_{in}'  - W_{out}' } \right)N - v\frac{{dN}}
{{dx}}
\]

The scattering events to which correspond the rates $W_{in,out}'$ are shown schematically in Fig. 1b). The last term describes the propagation of polaritons with a spatial velocity $v$, which is typically several tens of microns per picosecond in a waveguide, and which decreases as the energy approaches that of the exciton, because of the strong coupling. We have used an analytical approximation for the scattering rates based on the following simple phenomenological expression.

\begin{equation}
\Delta W = W_0 x_l \left( E \right)n_R^2 \left( {1 - \beta e^{ - \Delta /k_B T} } \right)
\end{equation}

where $\Delta W = W_{in}'  - W_{out}'$. This has allowed us to increase the efficiency of the simulations.

This expression takes into account the quadratic dependence of the exciton-exciton scattering rates on the reservoir density, the thermal distribution of excitons in the effective reservoir state, and the excitonic fraction of the polariton state, which depends on its energy. As can be seen in Fig. 2a), this analytical expression (dashed lines) fits very well the results of the numerical calculations for both materials (solid lines). Moreover, it allows the analytical solution of the equation (2), in order to find an estimate for the amplification factor:

\begin{equation}
\frac{{N\left( x \right)}}
{{N(0)}} \simeq e^{\frac{{\Delta W}}
{v}x}  - 1
\end{equation}

The results of the numerical simulation of the polariton propagation using equation \eqref{propag} are shown in Fig. 2b. We have considered a device size of $L$=80 $\mu$m. The signal amplification ratio is shown as a function of its initial intensity for GaN (black) and ZnO (red). In this region, $v_g\approx$25 $\mu$m/ps. We see that for a weak signal, an amplification of a factor of 10 can be obtained when operating at the optimal energy of about 35-40 meV below the exciton energy, depending on the material. Such a small amplifier size (as compared with e.g.\cite{Sugawara2004}, where a gain of 10 dB is obtained for 1 mm length) becomes possible because of the lower polariton velocity and high efficiency of the stimulated relaxation mechanism based on exciton-exciton scattering. The amplification is slightly higher for ZnO mainly because of the higher Rabi splitting, which allows a smaller group velocity for the same energy.

\begin{figure}[t]
\begin{center}
\includegraphics[width=0.99\linewidth]{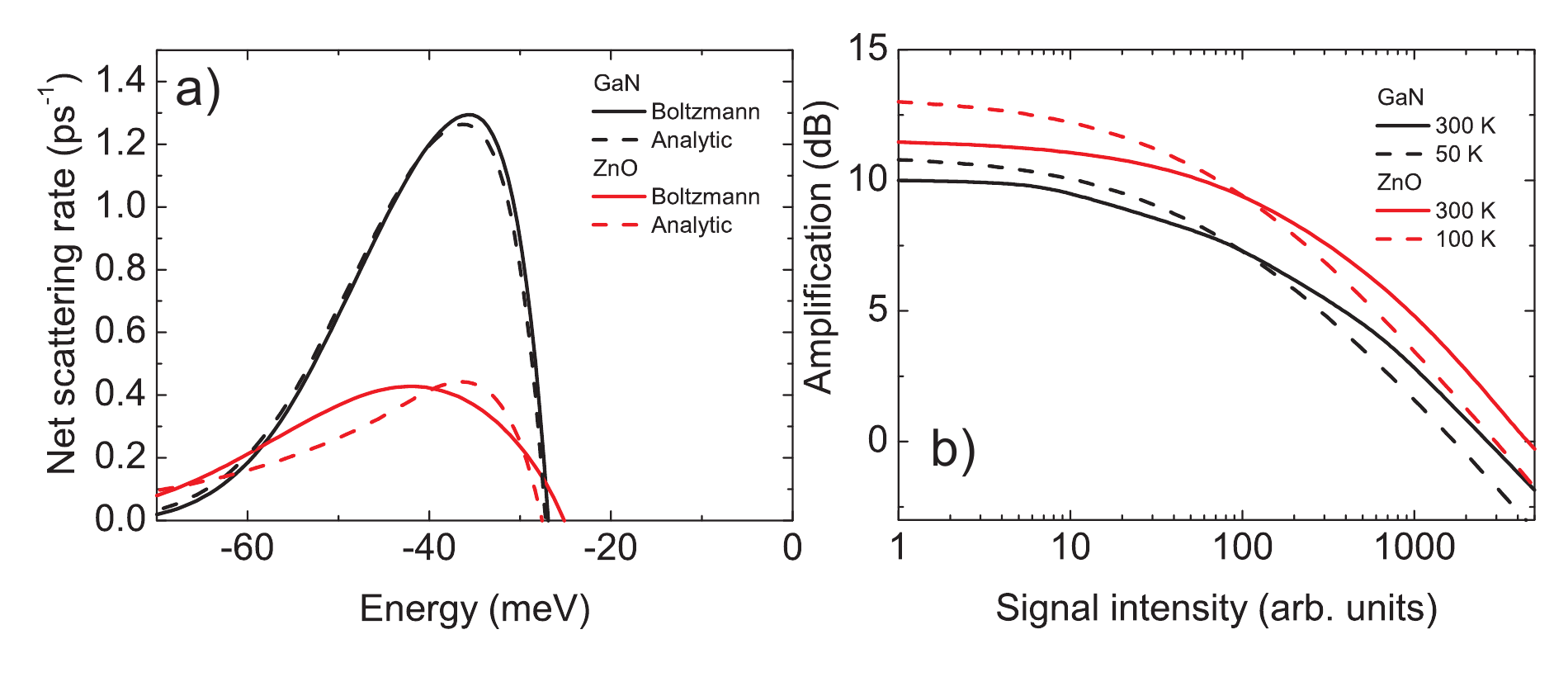}
\caption {\label{fig2} a) The scattering rates calculated using semi-classical Boltzmann equations (solid lines) and the analytical approximation (dashed lines) for GaN (black) and ZnO (red). b) The amplification ratio as a function of signal intensity for 80 $\mu$m-long device: GaN (black) and ZnO (red) at 300 K (solid lines) and optimal temperature of 50/100 K (dashed lines).}
\end{center}
\end{figure}

The amplification decreases for stronger signals, because of the important depletion of the reservoir. Indeed, only a fraction of excitons is available for rapid scattering into the signal. For a fixed pumping, the signal at the output cannot exceed a certain maximal value, no matter what is the initial signal intensity. The pumping cannot be increased above the threshold value, otherwise the amplifier passes into spontaneous lasing regime. The reservoir recovery time for the configuration considered in this calculation is $\tau_0$=10 ps (for small amplified signals), which allows a repetition rate of 100 GHz.

\begin{figure}[t]
\begin{center}
\includegraphics[width=0.8\linewidth]{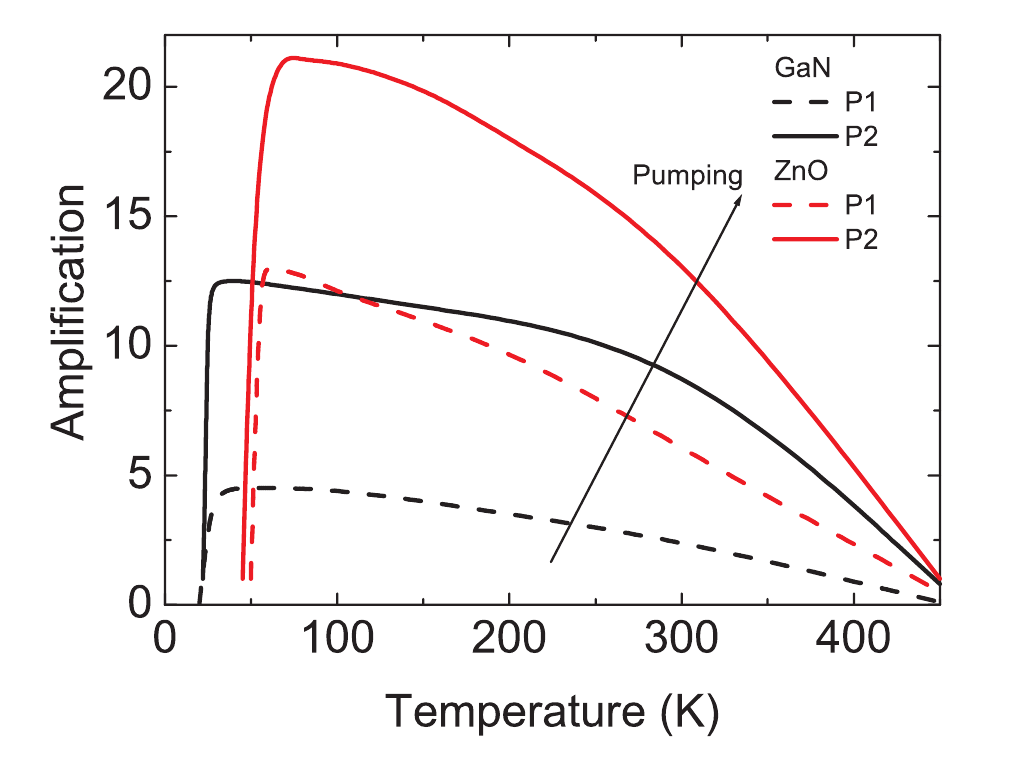}
\caption {\label{fig3} Amplification ratio as a function of temperature for GaN (black) and ZnO (red) amplifiers. Solid line - optimal pumping, dashed line - reduced pumping.}
\end{center}
\end{figure}

We have also studied the temperature behavior of the amplifier, shown in Fig. 3 for both GaN (black) and ZnO (red) active regions. At higher temperatures, the amplification ratio decreases for both materials. At lower temperatures, the cutoff is due to the onset of lasing at a frequency which is necessarily different from that of the test pulse. Stronger pumping allows to reduce the decrease of the amplification ratio at higher temperatures.

To conclude, we have demonstrated that the bulk GaN and ZnO waveguides in the strong coupling regime optically or electrically pumped just below the threshold of spontaneous lasing, can both operate as amplifiers of a micrometer size, providing an amplification factor of 10-20 and a repetition rate of up to 100 GHz. 
A simple waveguide structure should be enough for the relaxation to take place within the propagation time if GaN or ZnO are used as the active medium, but the efficiency can be enhanced using the slow light effects \cite{Chen2008}, with can be achieved by using photonic crystal slabs \cite{Chang-Hasnain2006}. For slow light regime, the propagation of interacting particles can take place in the superfluid regime, providing extra protection against loss mechanisms (backscattering, for example)  \cite{Leboeuf2010}. This system is also very promising for studying nonlinear effects in amplifiers.

We acknowledge support from the ANR Ganex (ANR-11-LABX-0014) project.

\end{document}